\documentclass[lettersize,journal]{IEEEtran}
\usepackage{amsmath,amsfonts}
\usepackage{algorithmic}
\usepackage{algorithm}
\usepackage{array}
\usepackage[caption=false,font=normalsize,labelfont=sf,textfont=sf]{subfig}
\usepackage{textcomp}
\usepackage{stfloats}
\usepackage{url}
\usepackage{verbatim}
\usepackage{graphicx}
\def\BibTeX{{\rm B\kern-.05em{\sc i\kern-.025em b}\kern-.08em
    T\kern-.1667em\lower.7ex\hbox{E}\kern-.125emX}}
\usepackage{balance}
\usepackage{xcolor} 
\usepackage{array} 
\DeclareMathOperator{\CProx}{CProx}
\DeclareMathOperator{\CProxNet}{CProxNet}

\begin{document}
\title{Diffusion Meets Unrolling: Compressive SAR Image Reconstruction with Interleaved Learned Corrections}
\author{Odysseas Pappas, Andrew C. M. Austin, Perla Mayo, Mohammad Golbabaee and Alin Achim
\thanks{This work was funded by DASA under the Defence Rapid Impact Open Call.

O. Pappas, P. Mayo and A. Achim are with VI Labs, School of Computer Science, University of Bristol. A.C.M. Austin is with the School of Electrical, Electronic, and Mechanical Engineering, University of Bristol. M. Golbabaee is with the School of Engineering Mathematics and Technology, University of Bristol.

Email: \{o.pappas, a.austin, pm15334, m.golbabaee, alin.achim\} @bristol.ac.uk. Corresponding Author: O. Pappas.
}}

\markboth{Journal of \LaTeX\ Class Files,~Vol.~18, No.~9, September~2020}%
{How to Use the IEEEtran \LaTeX \ Templates}

\maketitle

\begin{abstract}
Compressive Synthetic Aperture Radar (SAR) imaging, typically formulated as an inverse problem and solved with traditional iterative optimisation methods, can be very computationally expensive. We investigate the use of denoising diffusion probabilistic models (DDPMs) for compressive SAR image reconstruction, where the diffusion model is guided by a poor initial reconstruction from sub-sampled data obtained via standard imaging methods. We augment this data-driven method with model-driven interleaved refinement processes, inspired by traditional compressed sensing (CS) methods, to enhance sparsity and heavy tails in SAR images. Experimental results on real ERS-1 SAR datasets demonstrate consistent improvements over baseline DDPM and state-of-the-art methods, with notable gains in reconstruction fidelity and structural similarity, while incurring only modest computational overhead. The proposed hybrid framework effectively combines the representational power and efficiency of diffusion models with the interpretability and robustness of model-based optimisation, enabling accurate compressive SAR imaging.
\end{abstract}

\begin{IEEEkeywords}
Synthetic Aperture Radar, Compressive Imaging, Denoising Diffusion Probabilistic Models, Deep Unfolding.
\end{IEEEkeywords}

\section{Introduction}
\IEEEPARstart{S}{ynthetic} Aperture Radar (SAR) is one of the most capable and widely utilised remote sensing modalities. Its ability to produce high-resolution, detailed images of the Earth's surface without regard to weather conditions and the day/night cycle has led to its adoption in a number of applications. These include urban and agricultural land use monitoring, land surface deformation, flood detection and mapping, maritime and coastal traffic monitoring, disaster management and many others \cite{Tsokas2022}. 

SAR is a coherent radar imaging modality. The radar hardware, mounted typically on an airborne/spaceborne platform, transmits frequency-modulated signals to a scene on the ground and records the reflected radar echoes while in flight, achieving impressive spatial resolution due to the long synthetic aperture effect. The collected radar echoes (phase history) are not immediately suitable for human visual interpretation and must first be processed into formed images. Historically, SAR images have been formed using a number of classical computational imaging algorithms such as backprojection, the Range-Doppler algorithm, the Omega-k and the Range Migration algorithm \cite{Franceschetti1999}. Each of these offer distinct advantages and disadvantages in terms of speed, computational cost and image quality. 

Modern high-resolution SAR satellite platforms acquire a tremendous volume of raw phase history data that must be initially stored onboard the satellite and then transmitted to ground stations, processed into actionable image products, stored in datacentres and disseminated to end users. This situation gives rise to an obvious research question: is there a feasible way to produce SAR imagery from reduced sets of measurement samples, alleviating such overheads?

Compressed Sensing (CS) methods have provided the most promising answer to the above question \cite{Patel2010} \cite{Cetin2014} \cite{Kelly2012a}. CS provides a mathematical framework where under certain conditions SAR images can be reconstructed from parsimoniously sampled phase history data. Extensive research has been conducted in CS methods for SAR imaging \cite{Onhon2012} \cite{Uḡur2012}. CS-driven methods, however, have a number of shortcomings. The measurement matrix must satisfy certain mathematical properties which can make hardware implementations impractical, while iterative solving algorithms may require significant experimental tuning. Most importantly, as most CS SAR imaging relies on iterative optimisation algorithms to solve the reconstruction problem, the methods can become prohibitively computationally expensive. Very few papers in the literature address the application of such methods to realistically sized SAR scenes \cite{Yang2013} or tackle near-real-time processing \cite{Bi2020}.

With advances in machine learning and generative AI, a new generation of SAR imaging algorithms has started appearing. These include attempts to use neural networks to learn the complete mapping from raw phase history data to the image domain \cite{Rittenback2020}, without any subsampling or CS element considered. The use of neural networks poses unique challenges related to scene size as well as to the processing of complex-valued phase history data; a complex-valued CNN to process raw data into formed images is explored in \cite{Ji2024}.

Diffusion models have in turn come to be applied to SAR \cite{Wang2025}; methods for generating synthetic SAR scenes are presented in \cite{Qosja2024} and \cite{Crump2024}. Wang et al. \cite{Wang2024} propose a deep statistical framework for SAR imaging from actual raw data, where an imaging model is incorporated into the diffusion model and a measurement-consistency correction is applied at every step while sampling. This effectively amounts to a Bayesian plug-and-play formulation with the diffusion model as prior. While they note their method is robust to SAR data with missing pulses, the method is not strictly positioned as CS-like reconstruction of SAR images.

We propose here the use of a conditioned diffusion model for SAR CS imaging with an interleaved model-driven steering mechanism. Compressively sampled data are initially processed via standard SAR imaging algorithm, producing poor quality images that suffer from severe blur and aliasing artefacts. A Denoising Diffusion Probabilistic Model (DDPM) is jointly trained on matched pairs of good quality SAR images and corresponding poor quality reconstructions; the poor quality reconstructions serve as conditioning inputs to the DDPM, which learns to reconstruct a SAR image without the blur and aliasing present in them \cite{Pappas2025}. 

At inference, we interleave model-driven processes that refine the DDPM outputs, steering them closer to an optimal solution based on prior knowledge of the signal and encouraging desirable signal properties. These processes follow standard inverse problem formulations, and make use of regularisers that enhance sparsity and the heavy-tailed nature of the data, both characteristic of SAR images. To alleviate the need for experimental hand-tuning of optimisation parameters and to allow for dynamic variation of such parameters across the diffusion time schedule, we present learned network variants of these algorithms to be employed as interleaved processes. Our method combines the advantages of data-driven generative machine learning methods and model-driven classical optimisation methods, leveraging the representational power, robustness and speed of inference of machine learning methods with the interpretability and tractability of classical optimisation methods. 

Our contributions are specifically as follows. We demonstrate how interleaved model-driven processes aimed at enhancing sparsity and the presence of heavy tails can be incorporated into a DDPM-driven method for SAR imaging from compressively sampled sets of phase history data. We furthermore introduce learned, deep unfolded variants of these model-driven processes that remove the need for manual parameter tuning and dynamically adapt to the changing statistics of the signal of interest as the reverse diffusion process progresses. Finally, we demonstrate our proposed methods on a novel dataset of ERS-1 images formed from original phase history data and demonstrate how the interleaved processes enhance the performance of the baseline DDPM and produce images of quality comparable to the ground truth, outperforming other state of the art methods.

The remainder of our manuscript is structured as follows. Section II presents background information on related topics and techniques, namely a review of compressed sensing for SAR as well as an introduction to DDPMs. Section III discusses the DDPM compressive SAR imaging method of \cite{Pappas2025}, which serves as the backbone to our proposed approach. The proposed method in turn is introduced in Section IV, while Section V presents experimental results. Concluding remarks are finally presented in Section VI.


\section{Preliminaries}

\begin{figure*} [t]
   \begin{center}
   \begin{tabular}{c} 
   \includegraphics[width=0.925\textwidth]{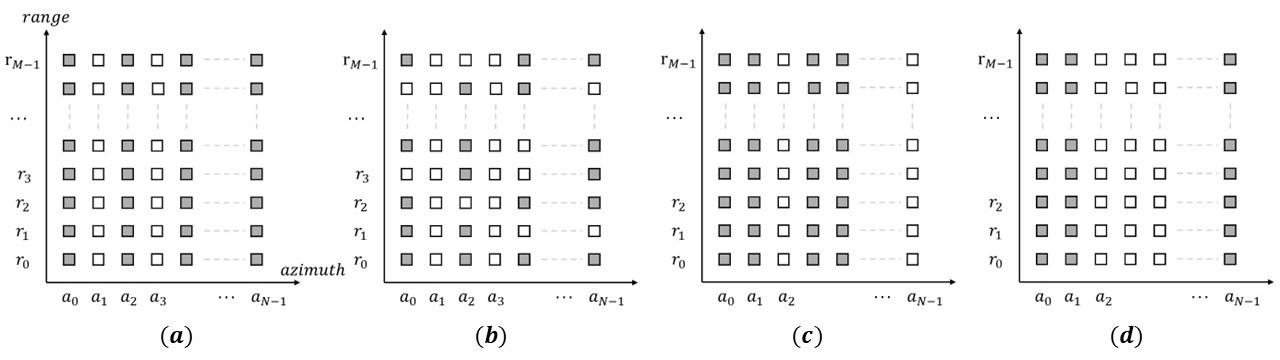}
   \end{tabular}
   \end{center}
   \caption[subsampl] 
   { \label{fig:subsamp} 
    Subsampling patterns for a SAR phase history matrix containing $M\times N$ samples. (a) Regular subsampling (at 50\%) along the azimuth. (b) Regular subsampling across azimuth with random subsampling across range. (c) Random subsampling across azimuth. (d) Aperture gap subsampling.}
\end{figure*} 

\subsection{Compressed Sensing for SAR}
Compressed Sensing (CS) is a sparsity-exploiting mathematical technique with which a signal can be successfully reconstructed from a parsimonious set of samples, typically numbering far fewer total samples than what the Nyquist theorem would require \cite{Donoho2006, Candes2008}. 

CS theory posits that if a signal of interest $x$ is sparsely representable in some domain, i.e. there exists a $K$-sparse signal $s$ that satisfies
\begin{equation}
\label{eq:CS1}
x = B s,~~~~~x \in \Re^N
\end{equation}
for some sparsifying basis $B$, then a reconstruction of $x$ from a set of measurements $y$ can be obtained via
\begin{equation}
\label{eq:CS2}
y = C x + n = C B s + n,~~~~~y \in \Re^M
\end{equation}
where $C  \in \Re^{M \times N}$ is the CS measurement matrix, with $M\ll N$, and $n$ denotes measurement noise. Typically, CS reconstruction methods rely on an inverse problem formulation whereby a minimisation problem of the form
\begin{equation}
\label{eq:inv1}
\hat{x} = \arg \min_x [F(x) = \Psi(y,\mathcal{A}x) + \psi(x)]
\end{equation}
is solved, with $F(x)$ being the cost function to be minimised, consisting of a data fidelity term $\Psi$ and a regularisation function $\psi$, and with $\mathcal{A}$ the forward operator.

Assuming independent and identically distributed Gaussian noise, the data fidelity term can be written as 
\begin{equation}
\label{eq:inv2}
\Psi(y,\mathcal{A}x) = \frac{||y-\mathcal{A}x||^2_2}{2\sigma^2}
\end{equation}
where$\sigma$ is the standard deviation of the noise. Assuming a prior probability density function (pdf) $p(x)$, the estimation of the signal of interest $x$ from the noisy observation $y$ is formulated as
\begin{equation}
\label{eq:inv3}
\hat{x} = \arg\min \left \{ \frac{||y-Ax||^2_2}{2\sigma^2} - \log p(x) \right \}
\end{equation}

The design of the CS measurement matrix $C$ must satisfy a number of properties in order for the reconstruction under CS theory to be valid. Namely, $C$ must satisfy the restricted isometry property (RIP) and be incoherent with respect to $\Psi$ \cite{Candes2008}. Random $C$ matrices can satisfy these properties but their use can be impractical in many applications. Therefore, the design of application-specific valid CS matrices still holds significant research interest \cite{Leier2014, Uḡur2012, Kelly2012a}. 

A practical pattern mentioned in SAR CS literature \cite{Leier2014} is regular undersampling along the azimuth direction, effectively discarding entire aperture positions (phase history columns) at regular intervals, as shown in Fig.~\ref{fig:subsamp}(a). While practical, this pattern does not actually satisfy the CS theoretical constraints; the regularity introduces coherent aliasing artifacts, and $C$ does not satisfy the restricted isometry property. Other patterns, shown in Fig.~\ref{fig:subsamp}, include random subsampling across range and aperture as well as aperture gaps. 

CS provides a way to form images from subsampled phase history data and thus opens up the possibility of significant savings in data acquisition, transmission and storage costs. A number of methods have been proposed in the literature, typically following an inverse problem formulation \cite{Cetin2014} and often incorporating phase correction (autofocusing) as a joint problem \cite{Onhon2012}. CS formulations additionally often make use of statistical models or regularisers (e.g. the alpha-stable distribution) that aim to promote sparsity or heavy tails in the data \cite{achim2010compressive, achim2015reconstruction}.



\subsection{Denoising Diffusion Probabilistic Models}

Denoising Diffusion Probabilistic Models (DDPMs) \cite{Ho2020} are a family of generative AI algorithms that rely on the modelling of a denoising process. Training images are corrupted by increasing levels of noise, as defined by a pre-set noise schedule, to the point where signals completely deteriorate into noise. During this forward diffusion process the model learns to predict the noise content of the image along this diffusion process; application of the reverse diffusion process amounts to a denoising process via which an image can be iteratively structured from pure noise.

Given $p(x_0)$ as the pdf of the training data, an uncorrupted input image $x_0$ becomes progressively corrupted via the addition of Gaussian noise into versions $x_1,x_2,...,x_T$ obtained according to:  
\begin{equation}
\label{eq:ddpmfwd}
p(x_t|x_{t-1}) = \mathcal{N} (x_t; \sqrt{1-\beta_t} \cdot x_{t-1}, \beta_t \cdot \textbf{I}),~~\forall t\in \{1,...,T\},
\end{equation}
where $T$ is the total number of diffusion steps, $\beta_1,...,\beta_T$ are hyperparameters representing the variance schedule, $\textbf{I}$ is an identity matrix of equal dimensions to the input image $x_0$ and $\mathcal{N}(x;\mu,\sigma)$ is the normal distribution of mean $\mu$ and variance $\sigma$ that produces $x$.

To sample at any arbitrary timestep $t$ we rewrite (\ref{eq:ddpmfwd}) using $\alpha_t=1-\beta_t$ and $\bar{\alpha}_t=\Pi_{s=1}^t\alpha_s$ as:
\begin{equation}
\label{eq:ddpmfwd2}
p(x_t|x_0) = \mathcal{N} (x_t; \sqrt{\alpha_t} \cdot x_0, (1 -\bar{\alpha}_t) \cdot \textbf{I})
\end{equation}
And therefore, given that the variances of the noise schedule are pre-set and not learnable, the sampling of $x_t$ (with diffusion noise $\epsilon$) can be denoted as:
\begin{equation}
\label{eq:ddpmfwd3}
x_t = \sqrt{\bar{\alpha}_t} \cdot x_0 + \sqrt{1-\bar{\alpha}_t\epsilon},~~~\epsilon~\mathcal{N}(0,\textbf{I})
\end{equation}
In applying the reverse process, a DDPM can generate new samples from $p(x_0)$ by beginning with a random Gaussian noise sample and iteratively forming it into a structured image by following the operation:
\begin{equation}
\label{eq:ddpmbwd}
p(x_{t-1}|x_t) = \mathcal{N}(x_{t-1};\mu_\theta(x_t,t),\Sigma_\theta(x_t,t)),
\end{equation}
where $\Sigma_\theta(x_t,t)=\sigma^2_t\textbf{I}$ is a fixed variance.

DDPMs can be used for unconstrained data generation but they can also be conditioned so as to guide the output to a desired product. The conditioning input can be a class label or an input signal such as a low-fidelity version of the desired output to be reconstructed, allowing DDPMs to be employed for a large number of computational imaging tasks such as image formation, in-painting and super-resolution \cite{Nichol2021, Dhariwal2021}.

Diffusion models have evolved significantly from the original DDPM architecture - from improvements to sampling (Improved DDPM, \cite{Nichol2021}), model conditioning (Guided DDPM, \cite{Dhariwal2021}) and tuning \cite{chen2023} to the move to latent-space representations \cite{Rombach2022} and the Stable Diffusion framework (the latter fall outside the scope of the present paper and are not discussed in more detail). Instead, we briefly introduce a number of methods that  directly modify the DDPM sampling process to add model guidance.

Diffusion Posterior Sampling (DPS) \cite{Chung2023} utilises a DDPM modified at inference to include a single data-consistency gradient step, approximating posterior sampling for inverse problems. An estimate of the clean image $\hat{x}_0$ is obtained and the term $p(y|\hat{x}_0)$ is used in the DDPM update equations in place of the conventional $p(y|x_t)$ formulation. 

Song et al. \cite{Song2021B} reformulated diffusion models in continuous time by expressing the forward noising process as a stochastic differential equation (SDE). Instead of learning a sequence of discrete reverse conditionals, the model learns a time-dependent score function and sampling is performed by numerically integrating the reverse-time SDE. This was used in conjunction with a Manifold Constrained Gradient (MCG) in \cite{Chung2022}, alternating between DDPM update steps and gradient steps on the data fidelity term $||y - \mathcal{A}x||^2_2$. 

DOLCE \cite{Liu2022} introduces model-driven refinements during sampling to a conditioned DDPM for compressive imaging. Unlike the single-step correction of DPS that is integrated into the update equations, DOLCE interleaves a complete optimisation routine. The DDPM produces a latent output at each timestep and DOLCE then solves a standard inverse problem of the form shown in Eq.~\ref{eq:inv3}. 
The onus here is on the data attachment term enforcing model consistency; the DDPM latent output is mapped back to the original measurement space via the physical model $\mathcal{A}$ and compared against a noised version of the original measurement data $y$. 
Note that none of the above-mentioned methods (DPS, MCG, DOLCE) have been applied to the specific problem of SAR imaging, and adaptation can be far from straightforward. 





\section{Diffusion-Based Compressive SAR Imaging}
\begin{figure} [t]
   \begin{center}
   \begin{tabular}{c} 
   \includegraphics[width=0.475\textwidth]{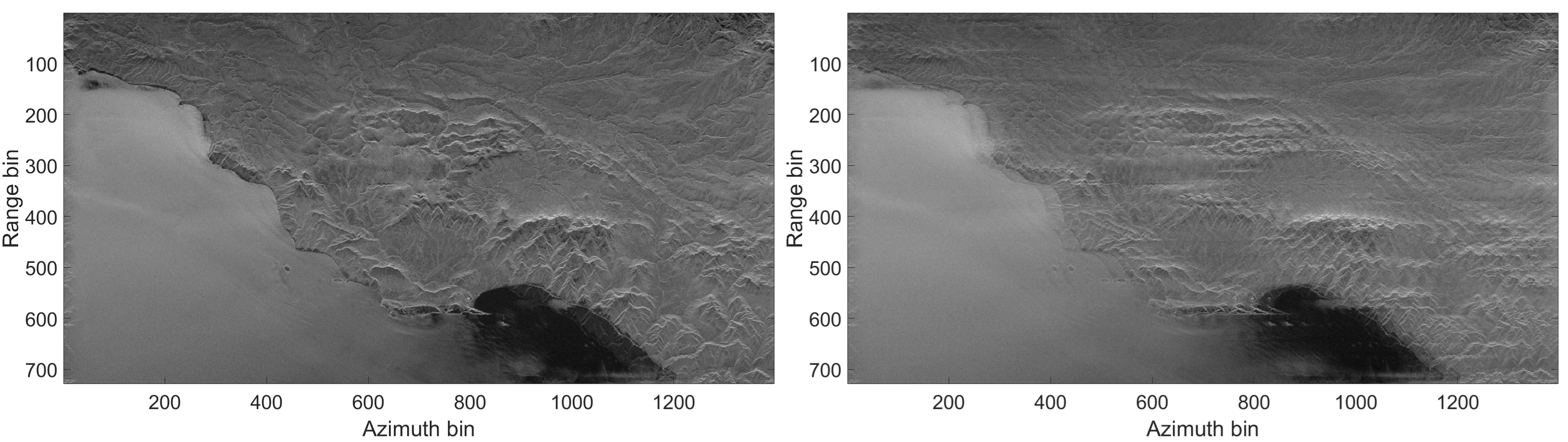}
   \end{tabular}
   \end{center}
\caption[subsampl] 
   { \label{fig:subsampexg} 
    Effect of phase history subsampling on classical reconstruction (via range migration) for ERS-1 images. Left: Reconstruction from fully sampled data. Right: Reconstruction from phase history data subsampled by a factor of 2 along the azimuth direction.}
\end{figure}

A number of DDPM-based methods have been proposed for SAR imaging and synthetic data generation \cite{Wang2024}; our earlier work \cite{Pappas2025} serves as the baseline for the proposed model-steered DDPM compressive SAR imaging framework. Training data consist of SAR images formed from fully sampled phase history data using conventional imaging algorithms (e.g. range migration) along with paired conditioning images that are poor quality, aliased reconstructions obtained from compressively sampled phase history data using the same imaging algorithms (Fig. \ref{fig:subsampexg}).

This allows the network, at inference time, to be presented with a poor-quality aliased SAR image produced from compressively sampled data and to effectively alleviate the aliasing and blurring artefacts within it, producing an output comparable to what could have been obtained from a fully sampled phase history. The structure of the CS measurement matrix affects the aliasing in the conditioning images, therefore training of the DDPM must be specific to a chosen subsampling pattern. Note this pattern would not need to satisfy RIP as we do not rely on conventional CS reconstruction theory.

DDPMs typically employ a U-Net backbone; scaling it to natively handle images at SAR resolutions is usually not computationally feasible. The processing of large images can instead be facilitated by patch-level processing \cite{Ozdenizci2023} or a move to latent-space methods \cite{Rombach2022}. We opt for the former as variational auto-encoders (VAEs) add significant training overheads and no suitable pre-trained VAE for SAR is known. During training, the data loader extracts a single patch from each training image, with its location randomly set; this allows the network to see multiple possible patches from each image as the training process iterates. At inference, the conditioning input image is segmented into overlapping tiles and these are processed independently through the network and then restructured into the original image dimensions.

\section{Model-Guided Diffusion for Compressive SAR Imaging}

\begin{figure*} [t]
   \begin{center}
   \begin{tabular}{c} 
   \includegraphics[height=3.85cm]{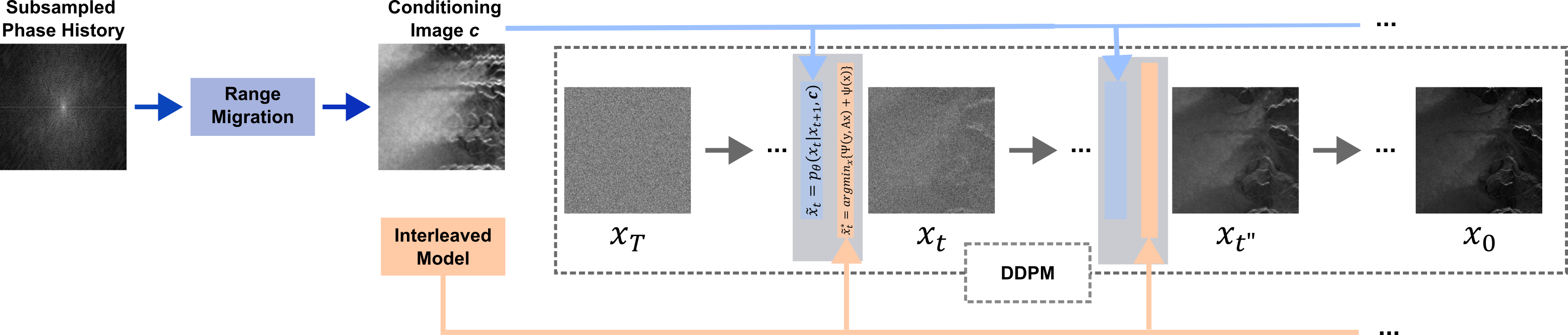}
   \end{tabular}
   \end{center}
   \caption[subsampl] 
   { \label{fig:interleavedddpm} 
Architecture overview of a DDPM for compressive SAR imaging with interleaved model refinement.}
\end{figure*}

We augment the performance of conditional DDPMs for compressive SAR imaging by interleaving a model-driven data refinement process, formulated as an optimisation problem, in between the inference steps of the DDPM. This architecture is illustrated in Fig.~\ref{fig:interleavedddpm}. 

\subsection{Design Considerations}

Prior approaches to DDPMs with interleaved model-driven techniques were discussed in~\cite{Chung2022} \cite{Liu2022}; a common thread is that the interleaved process takes the shape of an inverse problem optimisation rather than direct, implicit denoising or refinement of the data. This approach is crucial to allow for meaningful integration of physical models. 

\subsubsection{Data Attachment versus Regularisation}
The focus of the interleaved process has typically been on the data attachment term and how that can be shaped by the forward operator $\mathcal{A}$ embedded within the model. This views the DDPM itself as a strong learned prior, with the regularisation being implicit in the diffusion process. In methods like DPS and MCG no explicit regulariser is included in between the reverse diffusion timesteps; in DOLCE, regularisation is explicit  but it is driven by DDPM's score function serving as the regulariser. 

In our approach we interleave within the reverse diffusion steps a fully formulated optimisation process, but the aim here is not to refine the DDPM outputs with a data fidelity / forward operator model. We instead focus on the inclusion of explicit regularisation terms that enforce known physical properties of the underlying signal.
For SAR, such properties include sparsity and the presence of heavy tails. The DDPM itself acts as a prior, and will enforce these to the degree that they are learned from the training data; we here demonstrate that additional, explicit steering through a carefully chosen regularisation term can be beneficial to reconstruction performance.

\subsubsection{Choice of Refinement Signal}
In the context of DDPM-interleaved processing a couple of options arise as to what the observed data $y$ to be optimised are. The immediately obvious answer is to perform this interleaved optimisation process on the DDPM latent outputs $x_{t-1}$ (produced at timestep $t$) . As the inference progresses the interleaved optimisation helps to further steer $x_{t-1}$ towards the desired solution $x_0$ while the regularisation function encourages desirable properties. The updated latent variable $x_{t-1}^*$ is then fed back into the DDPM for the next inference timestep.

This approach poses a potential problem. The DDPM has been trained to follow a specific noise path, with interleaved processes only coming into effect at inference time in a plug-and-play fashion. Their effect on $x_t$ has not been accounted for during training; they can ``kick'' the DDPM off of the learned noise path and significantly deteriorate performance \cite{Chung2022}. It is possible to avoid catastrophic collapse of DDPM performance while still improving reconstruction by determining a ``subtle'' enough refinement ad hoc.



An alternative approach stems from the DDPM inference mechanism. Recall Eq.~\eqref{eq:ddpmfwd3}; at timestep $t$ the DDPM estimates the noise level $\hat\epsilon(x_t,t)$ and from  $\hat\epsilon$ and $x_t$ an estimate of the final denoised image $\hat{x}_0(x_t,t)$ is used to estimate $x_{t-1}$. Interleaved processing can therefore be performed on $\hat{x}_0$, which by definition is more representative of the underlying denoised SAR signal. The desirable effects encouraged by our regularisation mechanism (e.g. heavy tails) are applied directly on the underlying rather than the noisy signal, making the intervention more theoretically robust.

\begin{figure} [t]
   \begin{center}
   \begin{tabular}{c} 
   \includegraphics[width=0.45\textwidth]{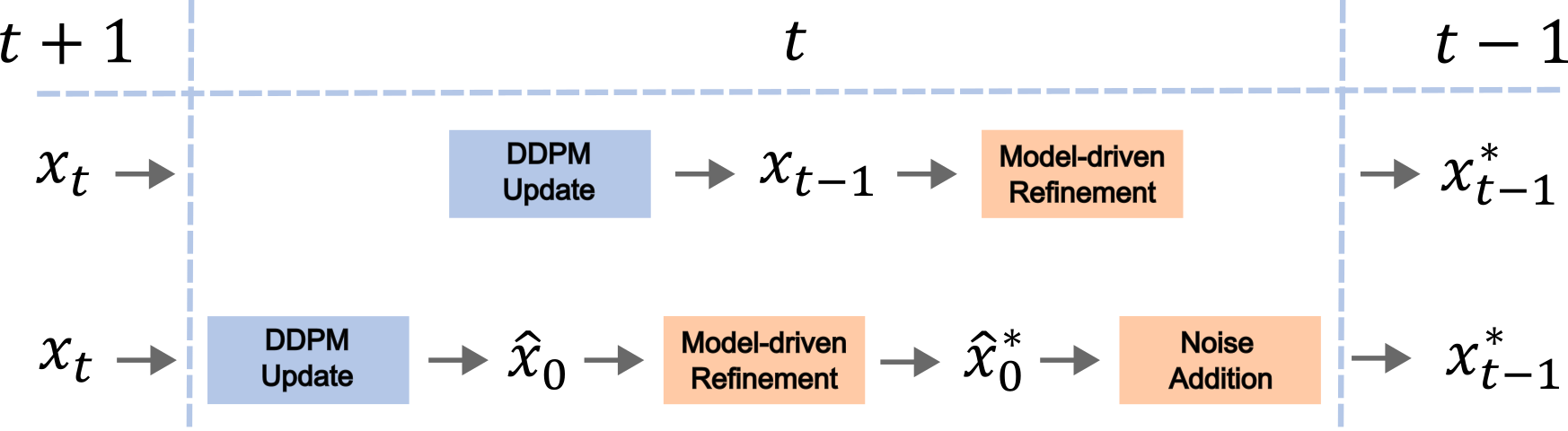}
   \end{tabular}
   \end{center}
   \caption[subsampl] 
   { \label{fig:xtx0} 
{Interleaved refinement schemes. Top path outlines direct refinement of the latent space noisy signal, bottom path shows refinement of underlying $\hat{x}_0$} signal.}
\end{figure}

The resulting $\hat{x}^*_0$ cannot be fed directly back to the DDPM to continue inference at $t-1$; noise must be added to bring the signal to the appropriate noise level, ``disguising'' $\hat{x}^*_0$ into $x_{t-1}^{*}$. This noise addition process is relatively trivial, and has the added benefit of eliminating the DDPM collapse challenge discussed above. Note that in this case the original DDPM latent output is discarded, replaced instead by the noised, refined $\hat{x}^*_0$. These schemes are shown in Fig. \ref{fig:xtx0}.

\subsubsection{Deep Unfolding for Interleaved Steering}
The use of iterative algorithms as interleaved model-steering processes poses a couple of challenges. They require careful manual tuning of parameters governing step size and regularisation strength, a task that can be further complicated if these parameters need to dynamically adapt across the reverse diffusion timescale. A large number of iterations is also typically required for convergence, and the computational burden of such algorithms being iterated to convergence after every diffusion timestep can become significant. These issues can be alleviated with the use of deep unfolded networks.

Deep unfolding \cite{Gregor2010} (or deep unrolling \cite{Monga2021}) is a technique in which neural networks are constructed so as to specifically mimic the performance of iterative optimisation algorithms. A deep unfolded neural network is designed so that one layer of it mimics the algorithmic process involved in one iteration of the analytical algorithm, with the parameters governing the analytical algorithm now turned into parameters learnable from training data. This approach combines the interpretability of traditional iterative algorithms with the learning power, speed and ease of use of deep neural networks. Unfolded networks also converge in far fewer layers than classical solvers require iterations, leading to lower computational cost, and better overall data efficiency. 


\subsection{Interleaved Methods}
As a first option for steering, we look at standard iterative algorithms for solving inverse problems, with regularisation functions aiming to enhance sparsity and heavy-tailed characteristics in the reconstructed image.

\subsubsection{FBS - ISTA}
Forward-Backward Splitting (FBS) \cite{Engl2000} refers to a general family of iterative proximal gradient descent algorithms for optimisation problems of the form:
\begin{equation}
\label{eq:fbs2}
\min_u \frac{1}{2}||\mathcal{A}u-y||^2_2 + \lambda ||u||_1
\end{equation}
where $u$ is the image to be reconstructed , $y$ are the observed data, $\mathcal{A}$ is the forward operator matrix, $\lambda$ is the parameter controlling regularisation strength and $||\cdot||_1$ is the $L_1$ norm used as the regularisation function. 


The forward gradient descent step yields intermediate variable $u$ as
\begin{equation}
\label{eq:fbsupd1}
u^{(i)} = x^{(i)} - \mu\mathcal{A}^T(\mathcal{A}x^{(i)}-y)
\end{equation}
where $i$ is the indexing variable and $\mu$ is the step size. The proximal operator of the $L_1$ norm amounts to a soft thresholding operation, and the backwards step becomes:
\begin{equation}
\label{eq:fbsupd2}
x^{(i+1)} = S_{\lambda}(u^{(i)}; \mu)
\end{equation}
where $S_{\lambda}$ is the soft thresholding operation with strength $\lambda$. This gives rise to the naming convention of Iterative Shrinkage Thresholding Algorithm (ISTA) for the standard $L_1$ regularised variant of FBS. The use of the $L_1$ norm enforces sparsity on the reconstructed data. Sparsity and SAR are often linked topics, with many SAR images being sparse in the spatial domain (due to isolated strong scatterers) as well as with sparsity being exploited in compressed sensing and sparse reconstruction formulations for imaging, autofocus, super-resolution and more \cite{Potter2010}. 

\subsubsection{Cauchy Proximal Splitting}
Cauchy Proximal Splitting (CPS) \cite{Karakus2020} is an adaptation of FBS that employs a novel Cauchy regulariser in place of the more commonly employed $L_1$. The Cauchy proximal operator has the effect of enforcing heavy-tailedness on the data, due to the heavy-tailed nature of the Cauchy distribution itself. This makes CPS particularly suited for applications such as SAR, and its effectiveness has been demonstrated on SAR-specific imaging, denoising and other inverse problems \cite{Karakus2020}.

The Cauchy distribution has a closed form probability density function given as
\begin{equation}
\label{eq:cauchypdf}
p(x) \propto \frac{\gamma}{\gamma^2 + x^2}
\end{equation}
where $\gamma$ is the Cauchy dispersion parameter controlling the spread of the distribution. From this, we formulate a Cauchy penalty function as 
\begin{equation}
\label{eq:cauchypenalty}
\psi(x) = - \log \left( \frac{\gamma}{\gamma^2 + x^2} \right) = \log(\gamma^2 + x^2) - \log(\gamma)
\end{equation}
which when making use of Eq.~\eqref{eq:inv3} and taking our penalty function $\psi(x)$ from Eq.~\eqref{eq:cauchypenalty} in place of $-\log p(x)$ leads to
\begin{equation}
\label{eq:cauchyestimate}
\hat{x}_{Cauchy} = \arg\min \left \{ \frac{||y-\mathcal{A}x||^2_2}{2\sigma^2} - \sum_{i,j}\log \left( \frac{\gamma}{\gamma^2 + x^2_{i,j}} \right) \right \}
\end{equation}

The Cauchy proximal operator can be derived \cite{Karakus2020} as the solution to the following cubic function:
\begin{equation}
\label{eq:cauchypoly}
u^3 - xu^2 + (\gamma^2 +2\mu)u - x\gamma^2 = 0
\end{equation}

The CPS algorithm, being a Cauchy-regularised variant of FBS, involves iteratively taking one gradient descent step on the data attachment term followed by the application of the Cauchy proximal operator (denoted $\CProx$). The forward step is as shown in Eq.~\eqref{eq:fbsupd1}, followed by.
\begin{equation}
\label{eq:cpsupd2}
x^{(i+1)} = \CProx(u^{(i)}; \gamma, \mu)
\end{equation}

\subsection{Learned Interleaved Methods}
The use of iterative algorithms such as ISTA and CPS poses a couple of challenges related to manual parameter tuning and adaptability to non-stationary inputs; we therefore propose the use of deep unfolding techniques to implement the previously described steering algorithms.

\subsubsection{LISTA}
ISTA is one of the many iterative algorithms amenable to deep unfolding, with its deep unfolded variants typically referred to as Learned ISTA (LISTA) \cite{Gregor2010}. LISTA converges much faster than ISTA, with typical implementations not going deeper than 5 to 7 iterations (as opposed to hundreds), and thus greatly alleviates the computational cost of the process being interleaved between inference steps.

The parameters governing step size and regularisation strength become learnable from the training data and, by making the LISTA network time-dependent (LISTA TD), these hyperparameters can evolve to adapt to the changing noise statistics as the reverse diffusion process progresses.


Initialisation is learned, with:
\begin{equation}
\label{eq:listaupd1}
z^{(0)} = S_{\lambda}(W_ex)
\end{equation}
and the update then continues as:
\begin{equation}
\label{eq:listaupd2}
z^{(i+1)} = S_{\lambda}(z^{(i)} + \mu W_tr^{(i)})
\end{equation}
\begin{equation}
\label{eq:listaupd3}
\hat{x} = W_dz^{(i_{max})}
\end{equation}
\begin{equation}
\label{eq:listaupd4}
r^{(i)} = x - W_d z^{(i)}
\end{equation}
with $z^{(i)}$ the latent variable at iteration $i$ and $r^{(i)}$ the residual term. $W_e$ is the learned encoder, $W_d$ is the learned decoder and $W_t$ is the transpose operator - these three effectively play the role of a learned forward operator and its transpose. Note $\lambda$ and $\mu$ are learnable scalars in the standard LISTA formulation. 

Time dependence is of particular interest here as typical LISTA formulations are trained on stationary data, and learn a set of hyperparameters to be used across any input data given. Stationarity is not a valid assumption in our case; the interleaved LISTA process will be presented with data of varying degrees of noise at different points of the diffusion process. LISTA will therefore learn hyper-parameters suitable for an average noise landscape, which can be problematic as it would lead to over-correction of the data in later stages of the inference process. One way to mitigate this is to impose an early termination criterion, where e.g. the LISTA process is interleaved only for the earlier part of the inference process where the DDPM outputs are at their noisiest.

A better way to address this issue involves embedding time dependence into the model, allowing it to natively adapt to the changing noise landscape via time-dependent parameters for threshold and step size $\lambda(\tau)$ and $\mu(\tau)$. We introduce a time dependence variable $\tau=t/(T-1)$ that represents a normalised reverse diffusion time index; this variable is passed to the model as a sinusoidal timestep embedding. 

\begin{equation}
\lambda(\tau)=\lambda_0+h_\lambda(e(\tau))
\end{equation}
\begin{equation}
\label{eq:listatd_mu}
\mu(\tau)=\mu_{\min}
+\left(\mu_{\max}-\mu_{\min}\right)
\sigma\!\left(h_\mu(e(\tau))\right)
\end{equation}
\begin{equation}
e(\tau)=g_\phi\!\left(\mathrm{PE}(\tau)\right)
\end{equation}
where $\tau=t/(T-1)$ is the normalised diffusion timestep, $\mathrm{PE}(\cdot)$ denotes the sinusoidal positional embedding, $g_\phi(\cdot)$ is a learnable multilayer perceptron, $h_\lambda(\cdot)$ and $h_\mu(\cdot)$ are learnable prediction heads, and $\sigma(\cdot)$ denotes the sigmoid activation.

The LISTA TD updates can therefore be written as
\begin{equation}
z^{(0)}=\mathcal{S}_{\lambda(\tau)}\!\left(W_e x\right)
\end{equation}
\begin{equation}
z^{(i+1)}=\mathcal{S}_{\lambda(\tau)}\!\left(z^{(i)}+\mu(\tau)W_t r^{(i)}\right)
\end{equation}
with Eq.~(\ref{eq:listaupd3}) and (\ref{eq:listaupd4}) holding as before.

LISTA outputs are not normalised or clamped in any way, so as to allow the DDPM to observe data at each timestep at the expected evolving dynamic range. Normalisation is used only within the loss function calculation, as the ground truth data are always in the fixed $[-1,1]$ range.

As with classical ISTA, there is the option here to have LISTA applied on the $x_t$ signal or the $\hat{x}_0$ signals. In the case of learned interleaved processes we found that training using cached $x_t$ data was particularly problematic, as any $x_t$ data from early in the reverse diffusion process are nearly pure noise. In attempting to map noise to a structured image the network is driven to hallucinate, and the outputs fail to retain image structure. We therefore restrict LISTA to the $\hat{x}_0$ case.


\subsubsection{DUCPS}

In analogy to CPS being a Cauchy-regularised variant of ISTA, we extend the formulation to the deep unfolded setting and obtain a Cauchy-regularised variant of LISTA, termed Deep Unfolded Cauchy Proximal Splitting (DUCPS). 

In contrast to LISTA, which retains an explicit proximal operator while learning the linear operators of the forward model, DUCPS replaces the proximal step itself with a learned mapping implemented via a convolutional neural network. This leads to a 2-block architecture with two separate sub-nets corresponding to the forward and backward steps of the CPS algorithm~\cite{Yang2024}. The updates can be written in an analogous fashion to LISTA as:
\begin{equation}
\label{eq:ducpsupd1}
    z^{(i+1)}=W_1 x^{(i)}+W_2y
\end{equation}
\begin{equation}
\label{eq:ducpsupd2}
    x^{(i+1)}=\CProxNet_\theta(z^{(i+1)})
\end{equation}
\begin{equation}
\label{eq:ducpsupd3}
    \hat{x}=x^{(K)}
\end{equation}
where $W_1=(1-\mu)I$, $W_2=\mu I$, given that $\mathcal{A}=I$, and $\CProxNet_\theta(\cdot)$ denotes a learned Cauchy proximal operator parametrised by a CNN. The subnets are iteratively called $K$ times at inference, thus implicitly controlling the unrolling depth. The update process is also shown in Algorithm \ref{alg:ddpm_ducps}.  

For DUCPS TD, these equations become time-dependent and parametrised as 
\begin{equation}
\label{eq:ducpstd_w1}
    W_1(\tau) = \left(1-\mu(\tau)\right)I
\end{equation}
\begin{equation}
\label{eq:ducpstd_w2}
    W_2(\tau) = \mu(\tau)I
\end{equation}
with
\begin{equation}
\label{eq:ducpstd_mu}
    \mu(\tau)=\sigma\!\left(\mu_0 + g_\phi(\tau)\right)
\end{equation}
Note Eq. (\ref{eq:ducpstd_mu}) is much simpler than its LISTA TD equivalent (\ref{eq:listatd_mu}), afforded due to the expressive power of $\CProxNet_\theta$. The DUCPS TD updates can therefore be written as
\begin{equation}
\label{eq:ducpstd_upd1}
    z^{(i+1)} = W_1(\tau)x^{(i)} + W_2(\tau)y
\end{equation}
\begin{equation}
\label{eq:ducpstd_upd2}
    x^{(i+1)} = \CProxNet_{\theta}\!\left(z^{(i+1)};\tau\right)
\end{equation}




The CNN denoiser acting as the Cauchy proximal operator consists of 5 convolutional layers, with 4 ReLU  activations and 1 residual skip. The CNN is trained to correct outputs towards the ground truth $x_0$, with the primary training objective combining an $L_1$ reconstruction loss and a Cauchy regularisation term encouraging heavy tails.  
The reference term $y$ is set equal to the DDPM estimate $\hat{x}_0$ so that the forward step anchors each iterate to the initial prediction. Applying DUCPS to the $x_t$ signal is avoided for the same reasons as for LISTA.

\begin{algorithm}[t]
 \caption{Interleaved DUCPS Refinement}
 \label{alg:ddpm_ducps}
 \begin{algorithmic}[1]
 \renewcommand{\algorithmicrequire}{\textbf{Input:}}
 \REQUIRE $t$, $\hat{x}_0(t)$, $K$, $\theta$, $\mu$, $\bar{\alpha}_{t-1}$
\STATE Initialize: $x^{(0)}, y \leftarrow \hat{x}_0(t)$
\FOR{$i = 0$ to $K-1$}
    \STATE \textbf{fwd:} $z^{(i+1)} \leftarrow (1 - \mu)\, x^{(i)} + \mu\, y$
    \STATE \textbf{prox:} $x^{(i+1)} \leftarrow \CProxNet_\theta\!\left(z^{(i+1)} \right)$
\ENDFOR
\STATE \textbf{return:} $\hat{x}_0^{\mathrm{DUCPS}} \leftarrow x^{(K)}$
\STATE \textbf{re-noise:} $x_{t-1} \leftarrow \sqrt{\bar{\alpha}_{t-1}}\,\hat{x}_0^{\mathrm{DUCPS}} + \sqrt{1-\bar{\alpha}_{t-1}}\,\hat{\epsilon}_t$
\RETURN $x_{t-1}$
\end{algorithmic}
\end{algorithm}

\section{Experimental Results}
In the following section we present a qualitative and quantitative evaluation of our proposed framework, in all 8 variants discussed previously, namely FBS/ISTA $x_t$,  FBS/ISTA $\hat{x}_0$, CPS $x_t$, CPS $\hat{x}_0$, LISTA, LISTA TD, DUCPS and DUCPS TD. We compare and contrast these to the baseline SAR CS DDPM method of \cite{Pappas2025} as well as to the DOLCE compressive imaging framework \cite{Liu2022}. Experiments were conducted on a workstation with an AMD Ryzen Threadripper 7960X CPU, an NVIDIA GeForce RTX 5090 GPU and 128 GB of RAM.



Applying DOLCE to SAR imaging is challenging due to the need for repeated mappings between measurement and image domains. In SAR, the relationship between phase history and image space is inherently many-to-many, with returns from across the scene contributing to each pixel. This renders patch-based processing problematic, as associating a local image patch with a corresponding subset of measurements is not tractable.
Hence, we adapt DOLCE by operating in the complex single-look (SLC) image domain as an intermediate measurement space, refining DDPM outputs against SLC data rather than the original phase history. The resulting operator $\mathcal{A}$ captures the effects of anisotropic mapping and phase loss introduced by multi-looking, but does not constitute a full SAR forward model. This departure from the original DOLCE formulation reflects the practical constraints of SAR imaging and illustrates the limited design space available to integrate model-based diffusion approaches in this setting.



\begin{figure*} [t]
   \begin{center}
   \begin{tabular}{c} 
   \includegraphics[width=0.925\textwidth]{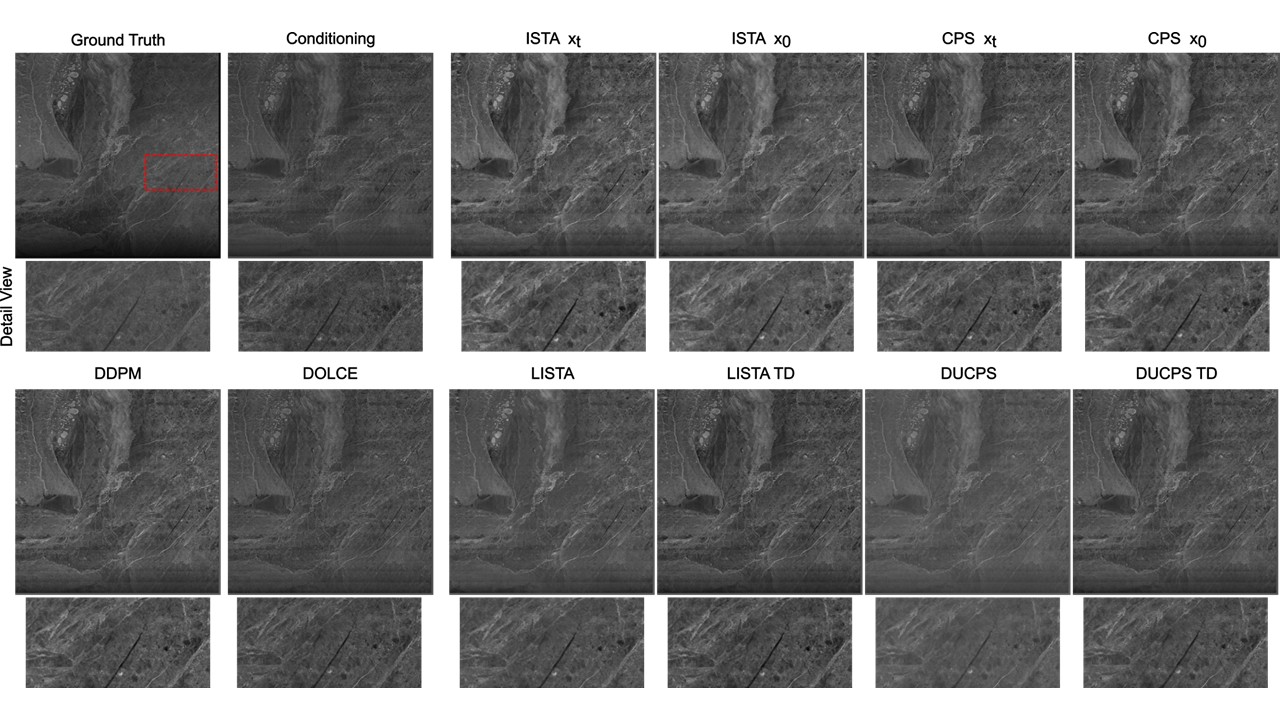}
   \end{tabular}
   \end{center}
   \caption[subsampl] 
   { \label{fig:res1} 
Experimental evaluation of proposed method (all 8 variants) on a full ERS-1 scene. Ground truth image, conditioning image, baseline DDPM and DOLCE reconstruction shown for comparison. Area in red dotted line shown in magnified view.}
\end{figure*} 

\subsection{Datasets}
For our experimental results, we have employed real SAR phase history data, decoded from Level-0 ERS products. Literature on SAR imaging, especially when employing machine learning algorithms, often relies on synthetic and simulated data, or on pseudo-phase history data generated from already formed SAR images. ERS-1 is a legacy SAR mission, but it remains one of the few platforms from which substantial volumes of usable Level-0 data can be obtained. 

Our training dataset consists of 500 Level-0 products acquired during January 1996 over North America, with additional images from the same period and location available for testing. Phase history data have been decoded from the Level-0 products and subsequently formed into SLC images using the range migration algorithm \cite{Franceschetti1999}. Additionally, the original phase history data were subsampled at a ratio of 50$\%$ regularly across the azimuth (as per Fig.~\ref{fig:subsamp}(a)), with noise floor values substituting the signal values in the original matrix columns corresponding to these non-sampled aperture positions. The subsampled phase histories were also processed via the range migration algorithm, producing aliased reconstructions with degraded quality, as shown in Fig.~\ref{fig:subsampexg}. 

The produced SLCs are then multi-looked (ML) with 20 looks in the azimuth and 4 in the range direction, and cropped to 1228$\times$1228 pixels (zero-padded as needed depending on patch size for network training). These matched pairs of fully-sampled and poor-quality CS-sampled ML reconstructions serve as the training data for our DDPM.

For testing, we utilise 5 ERS-1 products. Each ERS-1 scene is segmented into 256 overlapping tiles and these are independently processed. Quantitative assessment is performed via  standard reconstruction metrics such as Peak Signal-to-Noise Ratio (PSNR) and the Structural Similarity Index (SSIM). We also report the Learned Perceptual Image Patch Similarity (LPIPS) metric \cite{Zhang20128}, a state-of-the-art deep learning–based perceptual metric, while noting its uncertain suitability for SAR imagery. While full scene images (achieved by recompiling individual tiles) are presented for qualitative assessment, all quantitative evaluation is performed at tile-level. 

\subsection{Network Training Regime}

We utilise a Guided DDPM model \cite{Dhariwal2021} trained on matched pairs of full resolution ERS images and poor quality, aliased reconstructions. The network has been trained over 150,000 iterations, with a learning rate of $1e-4$, over 1200 diffusion steps and with a shortened linear noise schedule with $\beta_{end} = 0.01$, which is half the default DDPM $\beta_{end}$ value. The decision to shorten the noise schedule was motivated by experimental observations showing SAR images reached near complete diffusion well before the tail end of the process. 
 
The network is sized to handle image patches that are 256$\times$256 pixels and the conditioning image has been tiled with a stride of 64. The U-Net backbone of the DDPM has 192 base feature channels, 2 residual blocks per resolution level and attention layers at resolutions of 32$\times$32 and below.

LISTA and DUCPS were trained over 150,000 epochs, using an Adam optimiser with a learning rate of 1e-4 and an $L_1$  loss function (plus a Cauchy term for DUCPS). The ground truth data for LISTA's training are the same ground truth data as for the DDPM, i.e. SAR images formed from fully sampled phase history data using conventional methods.

A frozen copy of the above DDPM model was used to generate the on-the-fly cache of training data $\hat{x}_0$ utilised by LISTA and DUCPS. At every epoch, the LISTA/DUCPS training script extracts a random single patch from within the images, much like the DDPM training script. The DDPM is then called once per epoch to produce the required $\hat{x}_0$ data. 

\subsection{Experimental Evaluation}
For the interleaving with classical methods, we opt for a step size of $\mu = 1$ and a regularisation strength $\lambda = 0.001$ for ISTA (both $x_t$ and $\hat{x}_0$) whereas for CPS we set a step size of $\mu = 0.2$ and regularisation strength $\gamma = 50$. These parameters were experimentally derived to ensure DDPM collapse was avoided as well as to have approximately the same level of ``signal correction'' for both ISTA and CPS. The latter was achieved by logging the cumulative total of the mean absolute difference (CMAD) between the pre- and post-interleaved refinement signals, and ensuring it was constant at $\approx 0.25$.

For the interleaving with learned methods manual tuning is no longer necessary. Non-time-dependent models (LISTA and DUCPS) exhibit the behaviour discussed in Section IV, where non-stationarity leads to overly strong corrections in later diffusion stages. In the absence of direct control over regularisation strength, this is mitigated via early stopping, disabling interleaving in the final 100 sampling steps. Time-dependent variants require no such intervention. 



\begin{figure*} [t!]
   \begin{center}
   \begin{tabular}{c} 
   \includegraphics[width=0.925\textwidth]{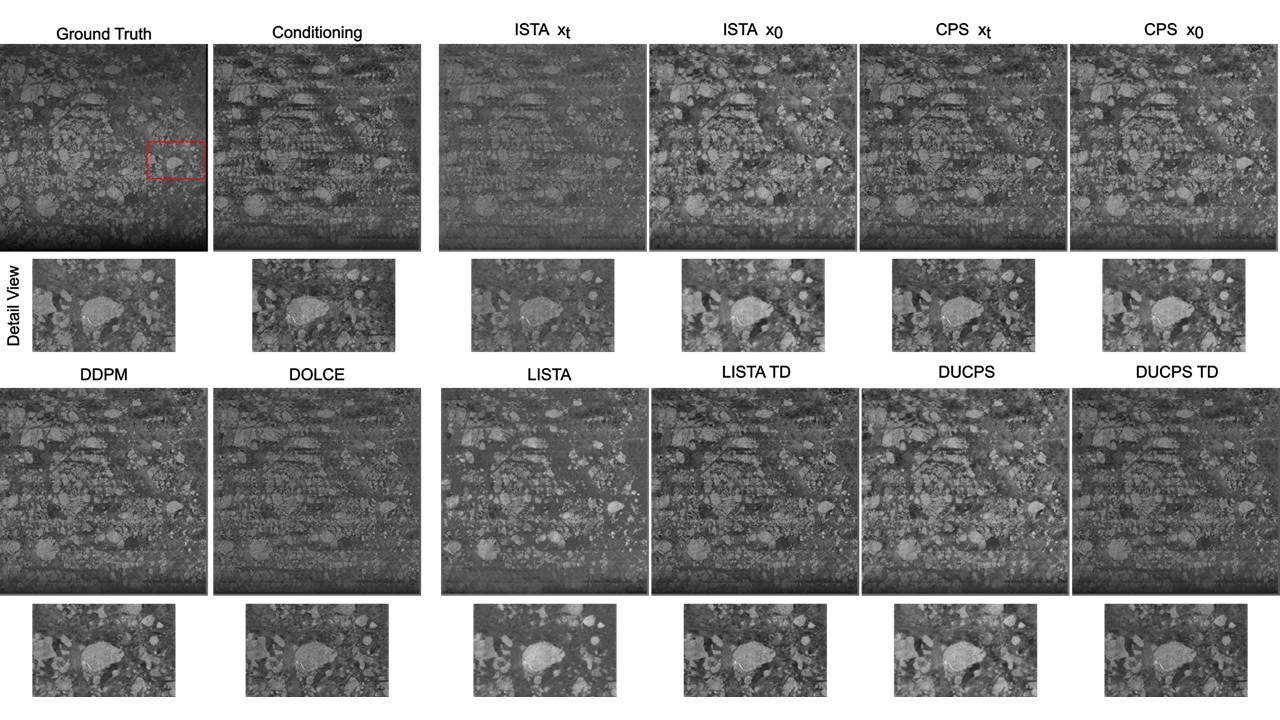}
   \end{tabular}
   \end{center}
   \caption[subsampl] 
   { \label{fig:res2} 
Experimental evaluation of proposed method (all 8 variants) on a full ERS-1 scene. Ground truth image, conditioning image, baseline DDPM and DOLCE reconstruction shown for comparison. Area in red dotted line shown in magnified view.}
\end{figure*}

Fig. \ref{fig:res1} and \ref{fig:res2} demonstrate the performance of our proposed method over two test scenes, with the 8 variants discussed, along with the ground truth image as well as the baseline DDPM method of \cite{Pappas2025} and DOLCE \cite{Liu2022}. From direct visual inspection it can be seen that all methods can produce images that are close to the original ground truth image, with the aliasing and blurring effects present in the conditioning image largely alleviated. 

The two classical refinement method, and especially ISTA, present relatively poor contrast performance when operating on the $x_t$ signal. The images appear much flatter in tone compared to both the ground truth as well as the output of the same algorithms when operating on $\hat{x}_0$. Considering that the baseline DDPM reconstruction does not appear as flat in terms of contrast, we hypothesize this is due to the DDPM performing unencumbered in the $\hat{x}_0$ domain, whereas direct modification of $x_t$ can deteriorate DDPM performance as discussed previously. This further supports $\hat{x}_0$ being the desired signal to refine on. 

It is worth noting that $\hat{x}_0$ produce sharp images with higher contrast than the conditioning input, and clearly enhance performance over the baseline DDPM (also quantitatively corroborated). Comparing ISTA $\hat{x}_0$ to CPS $\hat{x}_0$, the results of the latter appear sharper and with more fine detail, albeit the difference between the two is subtle. This is an indication that between sparsity and heavy tails the later is the more beneficial property to encourage in SAR images, though we cannot discount the potential of both being used in conjunction to enhance reconstruction.

LISTA presents relatively poor results when time dependence is not inherent to the model - even with an early stoppage criterion preventing it from being applied for the last 100 reverse diffusion sampling steps, the results show loss of fine detail. The LISTA TD variant, on the other hand, proves very capable, with the reconstructed images exhibiting good contrast and fine detail retention. For DUCPS we see an analogous situation, with a notable improvement in image quality on both variants. The non-time dependent variant is somewhat flat in contrast, but does not suffer from the same level of detail loss as LISTA. The DUCPS TD variant on the other hand shows the most promising performance of all proposed variants and baseline methods, with contrast closely matching the ground truth and conditioning input and sharp rendering of fine features. Certain fine features and areas of radiometric detail are more clearly discernible in the DUCPS TD reconstruction rather than in the original image. 

Finally, DOLCE produces images that are relatively flat in contrast albeit with satisfactory sharpness, and overall comparable to the baseline DDPM. The above evaluation points to DUCPS TD being the superior variant, clearly outperforming other methods and very importantly requiring no manual parameter tuning or early stoppage intervention.

\begin{table}[tbp]
\caption{Average Metric Scores \& Percentage Change over Baseline DDPM}
\centering
\scriptsize
\setlength{\tabcolsep}{5pt}
\newcolumntype{C}[1]{>{\centering\arraybackslash}m{#1}}
\begin{tabular}{|C{0.155\columnwidth}|C{0.08\columnwidth} C{0.08\columnwidth} C{0.08\columnwidth} | C{0.08\columnwidth} C{0.08\columnwidth} C{0.08\columnwidth}|}
\hline
\textbf{Method} & \textbf{PSNR} (dB) & \textbf{SSIM} & \textbf{LPIPS} & \textbf{PSNR ($\%$)} & \textbf{SSIM ($\%$)} & \textbf{LPIPS ($\%$)}      \\
\hline
\hline
SAR CS DDPM \cite{Pappas2025} & 11.62 & 0.35 & 0.48 & - & - & -\\
\hline 
\hline 
ISTA $x_t$ & 12.42 & 0.37 & 0.46 & +6.9 & +7.7 & +4.9\\
ISTA $\hat{x}_0$ & 12.59 & 0.35 & 0.51 & +8.6 & -0.4 & +4.6   \\
CPS $x_t$ & 11.76 & 0.36 & 0.47 & +0.9 & +5.2 & -3.5 \\
CPS $\hat{x}_0$ & 12.35 & 0.36 & 0.5 & +6 & +3.1 & +2.6 \\
\hline
LISTA & 11.03 & 0.35 & 0.55 & -4.5 & +5.8 & -2.3 \\
LISTA TD & 11.69 & 0.38 & \textbf{0.44} & +0.5& +9.1 & \textbf{-5.7} \\
DUCPS & \textbf{15.78} & 0.39 & 0.6 & \textbf{+36.5} & +18.1 & +12 \\
DUCPS TD & 15.68 &\textbf{0.43} & 0.5 & +34.8 & \textbf{+27.4} & +19.6 \\
\hline
\hline 
DOLCE \cite{Liu2022} & 11.72 & 0.39 & 0.53 & +0.7 & +15.3 & +11.9 \\
\hline
\end{tabular}
\label{tab:Results1}
\end{table}

Table \ref{tab:Results1} shows average quantitative results for the 5 test scenes available (two of which are seen in Fig. \ref{fig:res1} and \ref{fig:res2}). The reported average results are calculated on a tile-by-tile basis against the ground truth images and then averaged across all tiles and all scenes. We also report the score change, expressed as percentage, over the baseline DDPM.

The results show an overall trend of improvement in terms of PSNR for almost all methods over the stock DDPM reconstruction method of \cite{Pappas2025}. The only exception is LISTA, which scores low on all metrics, possibly due to excessive image smoothing/denoising despite the early stoppage. All other interleaved methods introduce an improvement in PSNR, with DUCPS variants outperforming all other methods by a notable margin. SSIM paints a similar picture, with all methods outperforming the baseline DDPM and DUCPS and DUCPS TD offering over 30$\%$ increase in PSNR.

LPIPS is less conclusive, as some methods appear to improve while others (such as DUCPS) present worse scores than the baseline method. However, given visual results seem to be corroborated by PSNR and SSIM, and taking into account LPIPS is trained on and optimised for natural images, we are inclined to attribute this behaviour to poor suitability of the metric to SAR. Note that for LPIPS lower values indicate improvement in performance. 


The experimental results demonstrate that the proposed model-assisted DDPM framework consistently enhances compressive SAR reconstruction compared to the baseline DDPM method of \cite{Pappas2025} as well as competing methods \cite{Liu2022}. While classical interleaved refinements such as ISTA and CPS provide modest and stable improvements in PSNR and SSIM, the deep-unfolded variants, particularly DUCPS and its time-dependent extension, deliver the most significant gains, achieving up to $36\%$ improvement in PSNR along with superior structural preservation and visual quality. Time-dependent learned methods further improve robustness by adapting to the evolving noise characteristics, avoiding issues such as over-smoothing observed in static models. These quantitative improvements are supported by qualitative results showing reduced aliasing and sharper feature recovery.


\subsection{Computational Overheads}

The inclusion of a model-driven interleaved refinement process obviously introduces new computational overheads to standard DDPM reconstruction. We have evaluated these overheads by measuring runtime (averaged over all tiles in a scene), and also provide an estimation of compute in GFLOPs as shown in Table \ref{tab:Compute}. Times are reported in seconds, and GFLOPs estimates are per tile per forward pass. The DDPM backbone has been estimated at 660 GFLOPs (constant for all methods) and the runtime has been measured at $\approx 5.6$~s, being practically constant for all methods as expected. Note that for the $\hat{x}_0$ variants there is an additional minor overhead ($2-3$~ms) for the noise addition step. 


\begin{table}[tbp]
\caption{Computation Times}
\centering
\scriptsize
\setlength{\tabcolsep}{6pt}
\newcolumntype{C}[1]{>{\centering\arraybackslash}m{#1}}
\begin{tabular}{|C{0.15\columnwidth}| C{0.12\columnwidth} C{0.12\columnwidth} C{0.125\columnwidth} C{0.125\columnwidth}|}
\hline
\textbf{Method} & \textbf{Total Time (s)} & \textbf{DDPM Time (s)} & \textbf{Overhead Time (s)} & \textbf{Overhead Compute (GFLOP)} \\
\hline
\hline
DDPM \cite{Pappas2025} & 5.6 & 5.6 & - & -\\
\hline
\hline
ISTA $x_t$ & 5.68 & 5.61 & 0.07 & 0.07 \\
CPS $x_t$ & 7.23 & 5.61 & 1.62 & 0.55\\
LISTA & 5.95 & 5.6 & 0.32 & 4.07 \\
LISTA TD & 6.02 & 5.59 & 0.4 & 4.09 \\
DUCPS & 6.65 & 5.59 & 1.1 & 105\\
DUCPS TD & 6.7 & 5.59 & 1.9 & 71 \\
\hline
\hline 
DOLCE \cite{Liu2022} & 7.53 & 5.58 & 1.85 & 1.7 \\
\hline
\end{tabular}
\label{tab:Compute}
\end{table}

As evidenced by the Table \ref{tab:Compute}, the interleaved model-driven processes do increase the overall computational time, but in all cases the introduced overhead is relatively minor, with the bulk of the processing time being the DDPM itself. This holds even for computationally complex methods such as DUCPS, which is in fact faster than its analytical counterpart.



\subsection{Compression Ratios}


\begin{figure*} [t!]
   \begin{center}
   \begin{tabular}{c} 
   \includegraphics[width=0.925\textwidth]{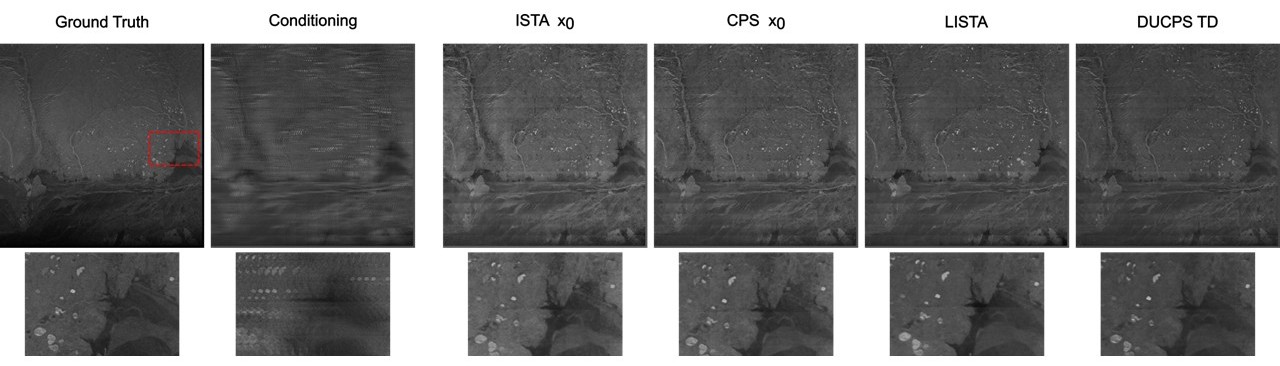}
   \end{tabular}
   \end{center}
   \caption[subsampl] 
   { \label{fig:5to1} 
   Experimental evaluation of proposed method (ISTA $\hat{x}_0$, CPS $\hat{x}_0$, LISTA, DUCPS TD ) on a full ERS-1 scene, with subsampling at $5:1$ ratio. Ground truth image and conditioning image shown for comparison. Area in red dotted line shown in magnified view.}
\end{figure*} 

In the above experiments, we utilised a compressive sampling matrix whereby alternate aperture positions were dropped, i.e., we employed a 50$\%$, or $2:1$, compression ratio over the raw phase history data. The proposed method can however extend to stronger compression ratios or different subsampling patterns. We present sample results using a $5:1$ compression ratio, discarding 4 out of every 5 apertures.

Heavier subsampling of the phase history data results in conditioning images that exhibit more severe blur and aliasing artefacts than in the example of Fig. \ref{fig:subsampexg}. This means that the DDPM must be re-trained on this new set of more severely deteriorated conditioning image, but once that is done the method can produce fine quality images even from these more substantially subsampled sets of phase history data.


\begin{table}[tbp]
\caption{Metric Scores for 5:1 Experiment}
\centering
\scriptsize
\setlength{\tabcolsep}{5pt}
\newcolumntype{C}[1]{>{\centering\arraybackslash}m{#1}}
\begin{tabular}{|C{0.28\columnwidth}|C{0.18\columnwidth}C{0.18\columnwidth}C{0.18\columnwidth}|}
\hline
\textbf{Method} & \textbf{PSNR (dB)} & \textbf{SSIM} & \textbf{LPIPS} \\
\hline
\hline
SAR CS DDPM \cite{Pappas2025} & 12.12 & 0.33 & 0.47 \\
\hline 
\hline 
ISTA $\hat{x}_0$ & 13.82 & 0.26 & 0.57 \\
CPS $\hat{x}_0$ & 12.82 & 0.33 & 0.48 \\
LISTA & 11.94 & 0.32 & 0.48 \\
DUCPS TD & \textbf{20.42} & \textbf{0.5} & \textbf{0.37} \\
\hline
\end{tabular}
\label{tab:Results3}
\end{table}

Fig. \ref{fig:5to1} shows experimental results of selected method variants, with a quantitative evaluation of this scene presented in Table \ref{tab:Results3}. The reconstructions are not quite as sharp as the ground truth and some blurriness can be noticed in comparison to the $2:1$ experiments, but the difference is not substantial. 

Quantitative results show a trend very much in line with the results reported previously, with all methods beside LISTA showing an improvement (over the baseline DDPM) in terms of PSNR. DUCPS TD scores the highest in terms of both PSNR, SSIM and LPIPS. 

The amount of aliasing induced in the conditioning image depends not only on the absolute compression ratio applied but also the subsampling pattern, characteristics of the SAR platform, as well as the multi-looking factors applied during image formation. There is therefore limited interest in determining the exact maximum compression ratio that can be handled by our method on this ERS-1 dataset, as that compression ratio would not necessarily translate to applications of the method on data from other satellites, with conditioning images formed via different imaging algorithms and with different multi-look ratios. However, the fact that the method does perform at similar level even at the heavier $5:1$ ratio is very much encouraging about the algorithm’s ability to deal with heavily subsampled phase histories.

\subsection{Effects of excessive regularisation strength}

To parametrise the ISTA and CPS algorithms, we first determine an appropriate step size $\mu$ by setting the regularisation strength weight to zero. This reduces both algorithms to just their forward step, i.e., a data fidelity term which when left to its own should accurately ``track'' the original data. We thus identified values of $\mu$ that produce an output that appears as close as possible to the original input while being as large as possible before oscillations occur. 

\begin{figure} [t]
   \begin{center}
   \begin{tabular}{c} 
   \includegraphics[width=0.45\textwidth]{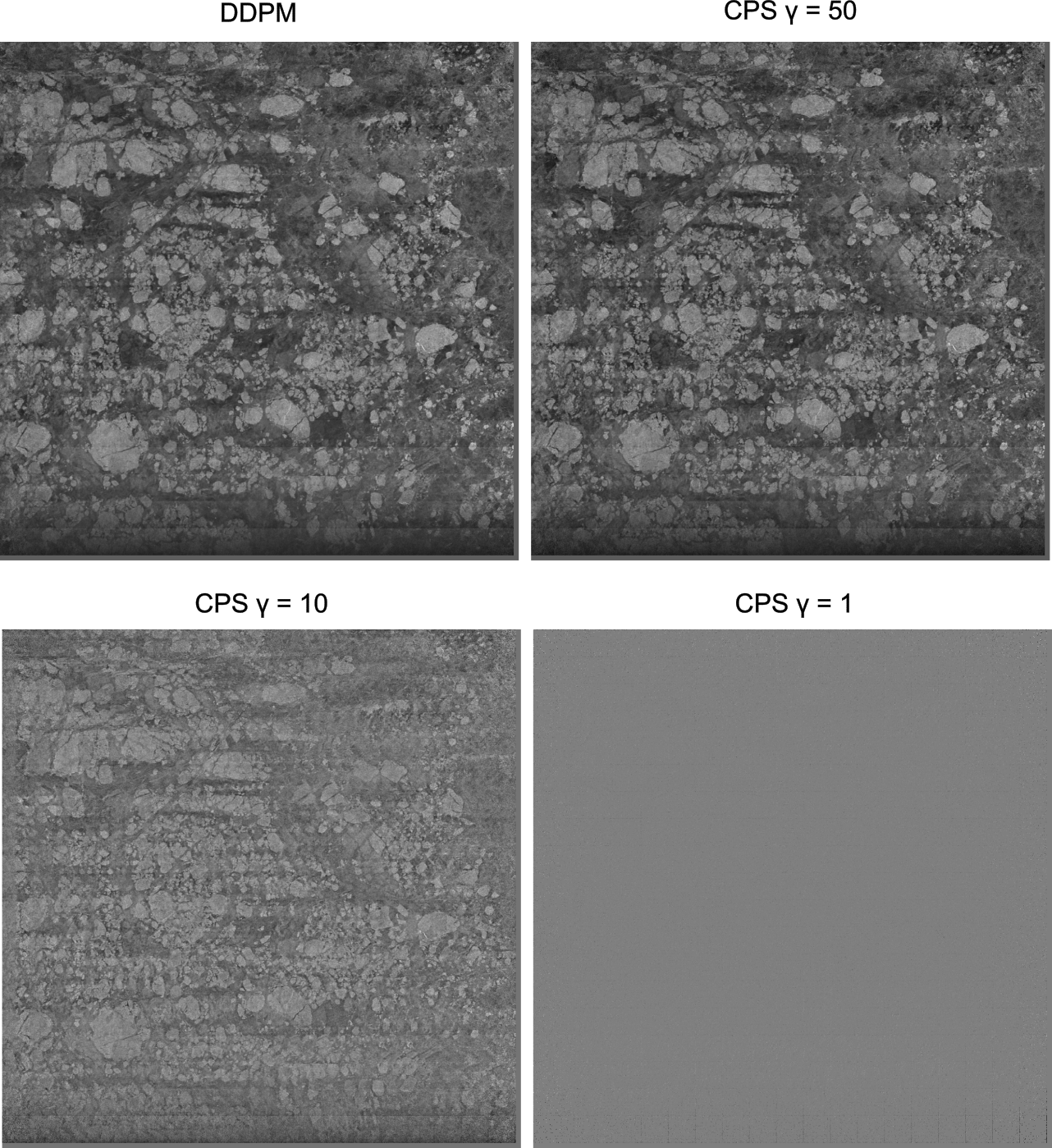}
   \end{tabular}
   \end{center}
   \caption[subsampl] 
   { \label{fig:collapse} 
    Deteriorative effect of excessive regularisation with the CPS $x_t$ method. Top left: Baseline DDPM with no interleaved process. Top right: CPS with $\gamma=50$. Bottom left: CPS with $\gamma=10$, loss of contrast and subpar DDPM performance. Bottom right: CPS with $\gamma=1$, complete collapse of DDPM output. Note that for CPS smaller values of $\gamma$ denote heavier regularisation weight.}
\end{figure}

With $\mu$ now set, we derive an appropriate regularisation weight experimentally. If regularisation becomes too strong the image starts exhibiting loss of contrast and fine detail due to excessive denoising/smoothing from the interleaved process. Past that point runs the risk of complete DDPM collapse as demonstrated in Fig. \ref{fig:collapse}.

The results shown in Fig. \ref{fig:collapse} signify the importance of manual parameter tuning for interleaved model-driven refinements. Similar deteriorative effects can be observed with non-time dependent learned methods, which have to be constrained via early stoppage in lieu of the lack of any direct control. In contrast, time-dependent learned methods require none such manual tuning, or early stoppage, and completely circumvent any potential DDPM collapse issues.

\section{Conclusions}
This paper introduced a model-assisted DDPM framework for compressive SAR imaging, combining diffusion-based generative modelling with deep-unfolded, model-driven refinement. A DDPM is conditioned on aliased reconstructions from subsampled phase history data, while interleaved unfolding steps explicitly steer the reconstruction toward SAR-specific distributions, promoting sparsity and heavy-tailed behaviour. By replacing manually tuned optimisation with learned, time-dependent unfolded modules, the framework adapts to the evolving signal statistics throughout the diffusion process.

The proposed model-assisted DDPM consistently improves reconstruction quality over the baseline, with most interleaved methods increasing PSNR and SSIM. Deep-unfolded approaches, particularly DUCPS and its time-dependent variant, achieve the strongest gains and best visual fidelity, while maintaining low computational overhead. These improvements persist even under stronger compression, demonstrating the robustness of the framework.

\balance

\bibliographystyle{IEEEtran}
\bibliography{cssar}



\end{document}